\documentclass[%
 reprint,
 superscriptaddress,
 amsmath,amssymb,
 aps,
 prl,
 floatfix,
]{revtex4-2}

\usepackage{graphicx}
\usepackage{dcolumn}
\usepackage{bm}
\usepackage{hyperref}

\begin{document}


\title{Training-Free Quantum Generative Paradigm via Local Parent Hamiltonians} 

\author{Shu Tian}
\affiliation{Department of Physics \& State Key Laboratory of Surface Physics, Fudan University, Shanghai 200438, China}

\author{Jiaqi Hu}
\affiliation{Center for Intelligent and Networked Systems, Tsinghua University, Beijing 100084, China}

\author{Rebing Wu}%
\affiliation{Center for Intelligent and Networked Systems, Tsinghua University, Beijing 100084, China}

\author{Yu Shi}%
\email{yu\_shi@ustc.edu.cn}
\affiliation{Wilczek Quantum Center, Shanghai Institute for Advanced Studies USTC, Shanghai 201315, China}
\affiliation{University of Science and Technology of China, Hefei 230026, China}
\date{\today}

\begin{abstract}

We propose a training-free quantum generative paradigm, which is fundamentally different from current generative models, which  demand substantial computational power, face practical scalability limits, and often function as opaque black boxes, despite their  remarkable success. We enable image and text generation without parameter training, by constructing a local parent Hamiltonian whose ground state encodes the target distribution and then solving the global Hamiltonian.  Rooted directly in quantum mechanical principles, this approach establishes a new pathway for generative modeling that leverages superposition and entanglement to maintain global consistency. 
\end{abstract}

\maketitle


\section{\label{sec:introduction}Introduction}

The rapid evolution of neural networks, particularly Transformer models, has led to remarkable breakthroughs in diverse generative tasks including image synthesis, machine translation, and natural language generation~\cite{VaswaniAshish2017Aiay, DevlinJacob2019BPod, DosovitskiyAlexey2020AIiW, GoodfellowIanJ.2014Gan, Sohl-DicksteinJascha2015Dulu, HoJonathan2020Ddpm, SongJiaming2021DDIM}. 
Despite these advances, classical generative models suffer from inherent limitations: the highly parameterized and  complex networks result in opaque, red{``black-box''} decision processes. Furthermore, scaling up these models exacerbates training instabilities, such as vanishing or exploding gradients and unstable optimization trajectories, often resulting in convergence failures or suboptimal solutions. A more fundamental challenge is their frequent inability to maintain rigorous long-range logical and contextual coherence~\cite{Shumailov2024AIModels,Manduchi2024ChallengesOpportunitiesGenAI,Hassija2024Interpreting,Dohmatob2024Tale}.
To address these challenges, we propose a quantum computing framework for generative modeling that departs fundamentally from classical neural networks by leveraging quantum properties rather than parameter tuning.

Classical computability is grounded in the Turing machine formalism, which describes the forward computational process: given a well-defined set of rules $M$ and an input $x$, compute the corresponding output $y = M(x)$~\cite{TuringA.M.1937OCNw}. In contrast, generative problems invert this paradigm: instead of computing outputs from known rules and inputs, they require reconstructing hidden generative rules $M$ or latent input patterns $x$  from observable outputs $y$, a problem known as inverse computation~\cite{Bennett1973Logical, Abramov2002Universal, Axelsen2011Reversible}. 

We argue that this inverse computational process has a natural mathematical formulation in quantum mechanics. It is well known that Quantum circuit models are equivalent to Turing machines in computability on one hand~\cite{DeutschDavid1985QTtC, AharonovDorit2008AQCI}, and  can be rigorously mapped to unitary transformations  through  quantum gates on the other~\cite{FeynmanRichardP.1982Spwc, LloydSeth1996UQS}. The unitary  transformations are governed by Hamiltonians while  the circuit's output is represented as the quantum state. This motivates us to  reframe the inverse computation for generative modeling as an inverse Hamiltonian reconstruction problem: given the ground states of a quantum system, which encode observed data patterns, recover the underlying Hamiltonian. This quantum  approach offers a valuable theoretical framework for generative problems.

Quantum computing naturally addresses generative challenges. The full output space is encoded as:
\begin{equation}
    |\Psi \rangle = \sum_i c_i |\text{Sequence}_i\rangle= \sum_i c_i \bigotimes_{k=1}^n |\text{token}_{ik}\rangle
\end{equation}
where $|c_i|^2$ corresponds to the probability of sampling the $i$-th sequence.
This captures three key quantum features, namely,  superposition, which enables parallel encoding of candidate tokens, entanglement, which enforces long-range correlations across positions,  and a physically grounded probabilistic sampling mechanism governed by $|c_i|^2$.

Existing quantum generative models primarily operate by quantizing classical neural architectures, such as Generative Adversarial Networks (GANs) and autoencoders. In this paradigm, models like Quantum GANs implement the generator and discriminator as parameterized quantum circuits, aiming to leverage quantum effects for more efficient sampling~\cite{LloydSeth2018QGAL, Dallaire-Demers2018Quantum, KhoshamanAmir2018Qva, Romero2017Quantum, Rudolph2024Trainability, Ma2025QuantumClassical}. Similarly, other frameworks like quantum autoencoders~\cite{Romero2017Quantum, Rao2023Learning} and quantum Boltzmann machines follow an analogous hybrid quantum-classical training approach~\cite{AminMohammadH, LiuJin-Guo2018Dloq, Zoufal2021Variational, Wilde2025Generative}. While these approaches represent valuable steps forward, they currently share similar challenges with their classical counterparts.
Accordingly, the pursuit of a practical and scalable quantum advantage through these hybrid paradigms constitutes an ongoing research endeavour.

In this Letter,
we introduce a novel quantum generative approach from the perspective of inverse computation. Departing from the prevalent paradigm of quantizing classical neural architectures and offering a distinct pathway to quantum generation,  our method is instead inspired by a classical algorithm called Wave Function Collapse  algorithm~\cite{MaximGuminWFC, KarthIsaac2017Wics, KarthIsaac2022WCGv}, which is actually unrelated to wave function collapse in quantum mechanics.
Our approach is quantum, mapping generation to finding the ground state of a quantum many-body Hamiltonian, where local patterns are encoded in overlapping local terms that naturally enforce compatibility constraints. The global ground state leverages entanglement for long-range consistency and is obtained via adiabatic, variational, or Grover algorithms.

We will first introduce 
our algorithm,
and  then apply it to image and text generations.  
A specific  example of three-letter words and 
the use of  Grover algorithm are demonstrated in the Appendices. 

\section{Quantum Generative Method 
\label{sec:qgm}} 


Our algorithm generates outputs that exhibit local similarity to the inputs. The basic requirement is \textit{pattern correspondence}---each contiguous segment in the output matches an identical pattern found in the input
. Building upon this, a stronger condition demands \textit{pattern density matching}: the frequency of each pattern in the output must also match that in the input.



\begin{figure}[htbp]
    \centering
    \includegraphics[width=\linewidth]{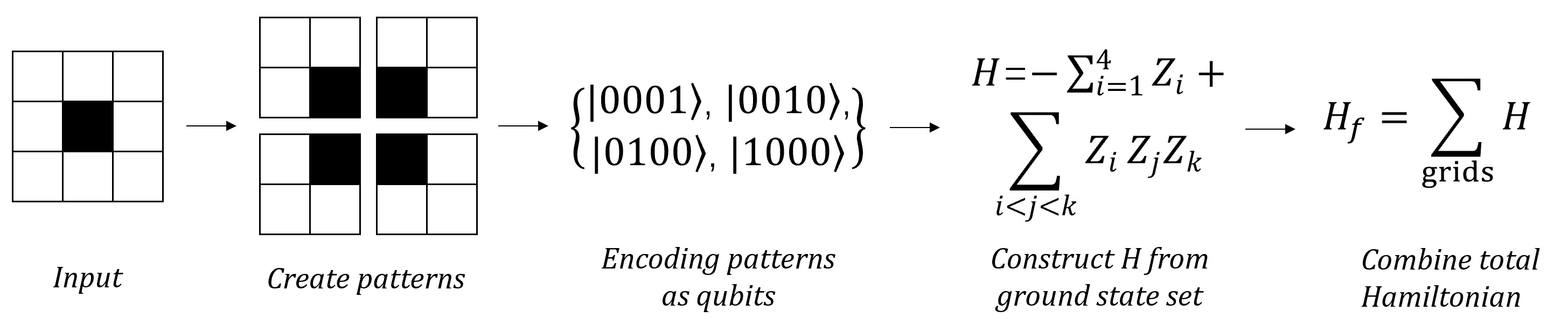}
    \caption{Flowchart of our generative model}
    \label{fig:flowchart}
\end{figure}

The core steps of our algorithm are  the following  (Fig.~\ref{fig:flowchart}). 

\begin{enumerate}
    \item Encoding Units: Map each atomic unit of the input (e.g., a pixel color or a word token) to a specific computational basis state of a qubit.
    \item Pattern Identification: Based on generation objective, patterns are either predefined, or screened out and cataloged from the encoded input state.
    \item Pattern State Construction: Map each unique pattern to a corresponding quantum state. The collection of all such pattern states forms a set $S$.
    \item Hamiltonian Formulation: Construct a local Hamiltonian $H$ whose ground states are exactly the states in set $S$.
    \item Total System Assembly: Combine local Hamiltonians to form a total Hamiltonian $H_f$, whose size corresponds to the desired output state.
    \item Ground State Solution: Solve for the ground state of $H_f$ and return this state as the output state.
\end{enumerate}

Different patterns intersecting at the same site impose multiple constraints. When $H_f$ reaches its ground state, each local subsystem adopts a valid pattern from set $S$, globally satisfying the local similarity condition. Pattern density matching is further achieved by encoding pattern frequencies via auxiliary weight qubits during the identification step.

\subsection{\label{Constructing Hamiltonian}Constructing Hamiltonian}

As detailed above, our Hamiltonian is a strict variant of the parent Hamiltonian problem~\cite{Affleck1987Rigorous, Perez-Garcia2007Matrix, Szehr2015Perturbation}: its ground-state manifold is exactly the span of $S = \{|\psi_1\rangle, |\psi_2\rangle, \dots\}$.

A direct and flexible construction is achieved via a spectral decomposition approach. We define the Hamiltonian as the sum of projectors onto all states in $S$:
\begin{equation}
    H = -E_g \sum_{i} | \psi_i \rangle \langle \psi_i |,
    \label{eq:spectral_hamiltonian}
\end{equation}
where $E_g > 0$ sets the energy scale. By construction, $| \psi_i \rangle$ is a ground state of $H$ with energy $-E_g$, while all states orthogonal to this subspace are degenerate excited states with energy $0$. This yields a finite energy gap $\Delta = E_g$ which assures adiabatic evolution or variational ground-state preparation.

To implement this Hamiltonian within a quantum circuit, we consider the unitary time-evolution operator $U(\theta) = e^{-i H \theta}$. Using Eq.~\eqref{eq:spectral_hamiltonian}, this becomes:
\begin{equation}
\begin{gathered}
    U(\theta) = e^{i\theta E_g\sum_i|\psi_i\rangle\langle\psi_i|} = \prod_i U_i, \\
    U_i = e^{-iE_g\theta|\psi_i\rangle\langle\psi_i|} = \mathrm{diag}(1,\dots,e^{-iE_g\theta},\dots,1).
\end{gathered}
\label{eq:unitary}
\end{equation}
Thus, 
$U(\theta)$ reduces to a series of controlled-phase gates.

The spectral method offers a transparent construction,
but imposes the specific requirement of zero-energy excited states—a sufficient but not necessary condition for a parent Hamiltonian. Alternative, more general constructions exist. For example, one can introduce ancillary  qubits to systematically map the constraint satisfaction problem onto the ground state of an Ising-type Hamiltonian using techniques such as linear programming~\cite{BianZhengbing2014Douq}. 

\section{\label{sec:graph generation}Graph Generation} 

\begin{figure}[ht]
    \centering
    \includegraphics[width=\linewidth]{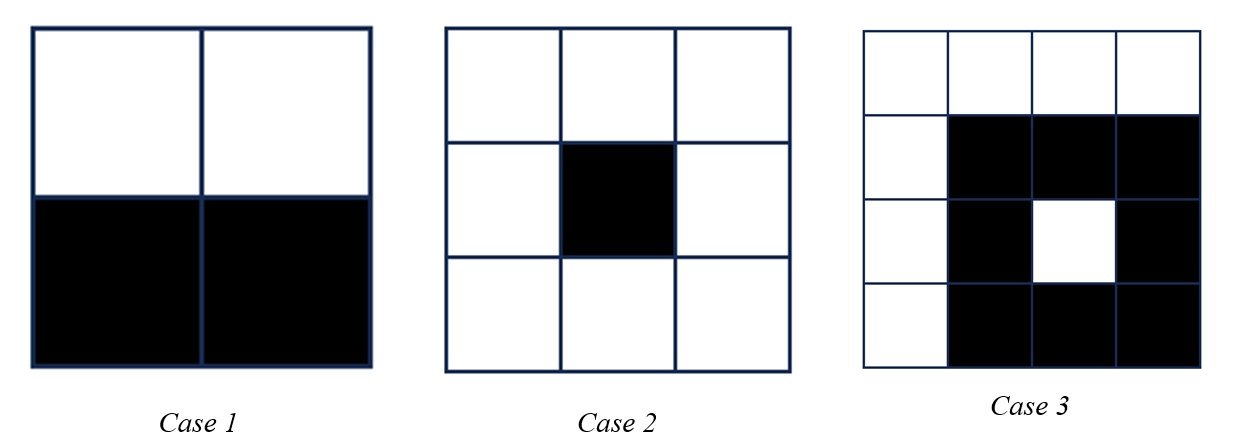}
    \caption{Three different original images of graph generation}\label{fig:case}
\end{figure}

Using the algorithm proposed above, we map each pixel to a qubit ($|0\rangle$ for white, $|1\rangle$ for black) and define patterns as all symmetry-equivalent $2\times 2$ patches from the input image Fig.~\ref{fig:case}, each encoded as a position-encoded state $(p_{\rm NW}, p_{\rm NE}, p_{\rm SW}, p_{\rm SE})$. Simulations are performed with 25 qubits on A800GPU using Quest~\cite{quest2019} and TensorCircuit~\cite{Zhang2023TensorCircuit}.
\begin{equation}
    |\psi_p\rangle = |p_{\text{NW}}\rangle_{\text{NW}} \\
    \otimes |p_{\text{NE}}\rangle_{\text{NE}} \\
    \otimes |p_{\text{SW}}\rangle_{\text{SW}} \\
    \otimes |p_{\text{SE}}\rangle_{\text{SE}}, 
\end{equation}
which is defined as the  pattern basis state corresponding to pattern $p$, i.e. the pattern basis set is  
$S = \left\{ |\psi_p\rangle \,\big|\, \forall \text{pattern p} \right\}$

\begin{figure}[ht]
    \centering
    \includegraphics[width=\linewidth]{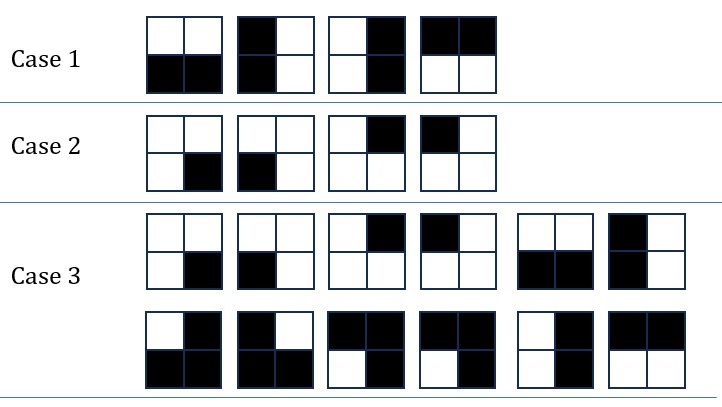}
    \caption{Patterns in each case}\label{fig:pattenrs in 2 cases}
\end{figure}
Then construct pattern-specific Hamiltonians with ground states corresponding to $S$:
\begin{equation}
    H = \begin{cases}
        Z_1 Z_4 + Z_2 Z_3 & \text{(Case 1)} \\
        -\sum\limits_{i=1}^4 Z_i + \sum\limits_{\substack{i<j<k \\ \text{3-body}}} Z_i Z_j Z_k & \text{(Case 2)} \\
        \sum\limits_{i\neq j} Z_i Z_j + Z_1 Z_2 Z_3 Z_4 & \text{(Case 3)}
    \end{cases}
\end{equation}

\begin{figure}[ht]
    \centering
    \includegraphics[width=\linewidth]{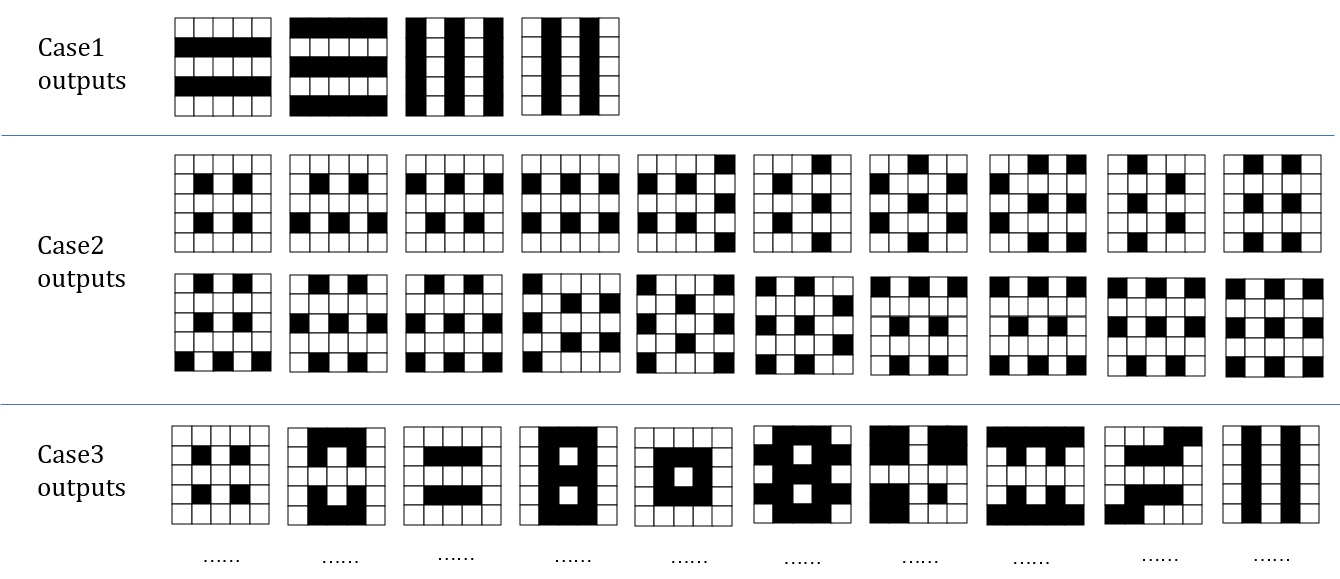}
    \caption{Each $5\times5$ image is associated with a product state $| \psi_i \rangle$. All generated states conform to the specifications of Case 2 and collectively comprise the complete set of $5\times5$ states satisfying the local similarity constraints.
    }\label{fig:output_graph}
\end{figure}

\begin{figure}[ht]
    \centering
    \includegraphics[width=\linewidth]{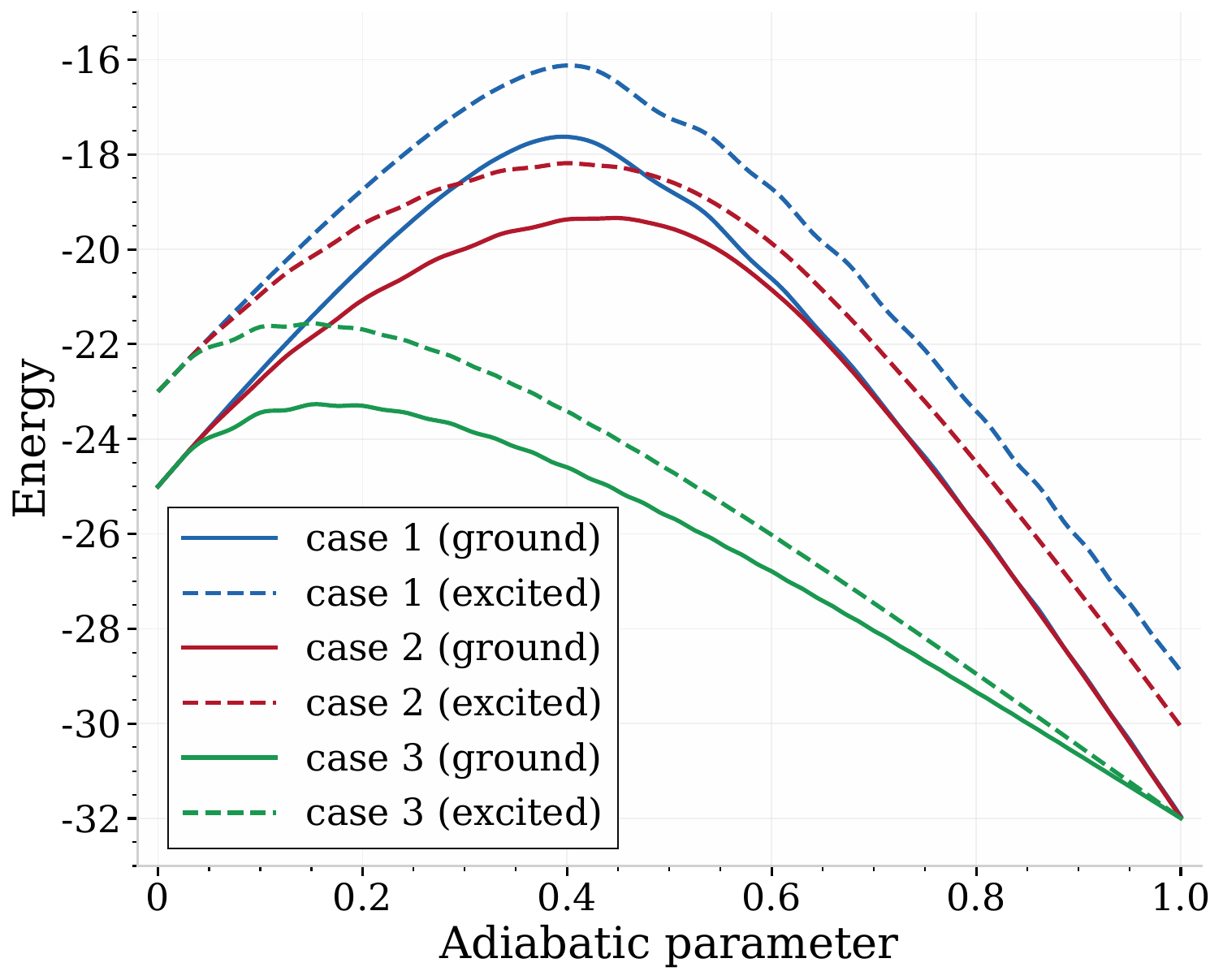}
    \caption{In the first two cases, the initial excited states evolve into final excited states while maintaining a finite energy gap throughout. The third case does not indicate adiabatic failure. Instead, even when starting from an excited state, the system can still converge to the ground state configuration after the evolution,
    }\label{fig:graph_ad}
\end{figure}

The global Hamiltonian for $N\times N$ image is:
\begin{equation}
    H_f = \sum_{2 \times 2 \text{ grids}} H
\end{equation}
The ground states of $H_f$ can be achieved through different quantum protocols, including:
\begin{itemize}
    \item Variational method: $|\psi_{\text{out}}\rangle = \arg\min_{\theta}\langle H_f\rangle_{\psi(\theta)}$

    \item Adiabatic evolution: $H(t) = (1-t/T)H_0 + (t/T)H_f$
\end{itemize}
Fig~\ref{fig:output_graph} illustrates the generated quantum states of the three cases.
The adiabatic protocol converges to a final state that is a coherent superposition of nearly all configurations satisfying the local similarity condition: 
\begin{equation}
| \psi_{\text{ad}} \rangle = \sum_{n} c_n | \psi_n \rangle + \epsilon_{\text{trotter}}, \| \epsilon_{\text{trotter}} \| < 0.01
\end{equation}
The evolution required only 200 time steps, which is two orders of magnitude fewer than typical many-body adiabatic protocols. The corresponding ground-state and first-excited-state energies under three representative conditions are shown in Fig.~\ref{fig:graph_ad}.
Variational quantum algorithms (VQAs) consistently converge to a random targeted single state or a simple superposition of very few states with high fidelity.


\section{\label{sec:text generation}Text Generation}
For text generation, our generative models can be regarded as a quantum n-gram model~\cite{Jurafsky2014Speech,Li2019CNM}. Following the procedure given above,
we first define a fixed vocabulary that maps each word to a computational basis product state, allowing $N$ qubits to represent up to $2^N$ distinct tokens.

For each $n$-gram whose joint probability $P(x_1, \dots, x_n)$ is non-zero, we perform a discretized weight encoding via a scaling and flooring operation:
\begin{equation}
    w = \lfloor C \cdot P(x_1, \dots, x_n) \rfloor,
    \label{eq:weight_encoding}
\end{equation}
where the constant $C = 2^k$ defines a $k$-bit resolution. The floor operator $\lfloor \cdot \rfloor$ discards $n$-gram patterns with probabilities smaller than $1/C$. We map pattern $\mathcal{P}$  to a set of $w$ orthogonal quantum states:
\begin{equation}
    | \psi^k_i \rangle = \bigotimes_{j=1}^n | \text{x}_j \rangle_{t_j} \otimes | k \rangle_{\text{weight}},
    \qquad k = 1,\dots,w,
    \label{eq:token_state_representation}
\end{equation}
where each of the first $n$ token registers $| \text{x}_j \rangle$ uses $m = \lceil \log_2 V \rceil$ qubits, with $V$ being the fixed vocabulary size of the n-gram model and the weight register $| k \rangle_{\text{weight}}$ is allocated $m' = \lceil \log_2 C \rceil$ qubits for storing the discretized weight.

For a text of \(T\) tokens, there are \(T-n+1\) overlapping \(n\)-grams. Each \(n\)-gram is assigned a local Hamiltonian term \(H_k\), the global Hamiltonian is constructed as the sum of these local terms:
\begin{equation}
H_f = \sum_{k=1}^{T-n+1} H_k .
\end{equation}

A candidate ground state of the full system can be expressed as:
\begin{equation}
    | \psi \rangle = \Bigl( \bigotimes_{j=1}^T | x_j \rangle_{\mathcal{R}_j} \Bigr)
    \otimes \Bigl( \bigotimes_{k=1}^{T-n+1} | v_k \rangle_{\mathcal{W}_k} \Bigr),
    \label{eq:full_ground_state}
\end{equation}
where the token registers $\mathcal{R}_j$ encode the sequence $(x_1,\dots,x_T)$, and the weight registers $\mathcal{W}_k$ store the n-gram statistics.
The probability of the ground-state configuration is proportional to the joint probability of all its constituent patterns, which satisfies:
\begin{equation}
    \prod_{k=1}^{T-n+1} w_k \;\propto\; P(x_1,\dots,x_T),
    \label{eq:weight_proportionality}
\end{equation}
with $w_k$ the quantized weight of the $k$-th n-gram.
The final state is a superposition of all possible outcomes, where the probability density of each output is proportional to its occurrence probability, i.e.,
the maximal weight state $\arg\max_{\psi} \sum w_k$ corresponds to the minimal perplexity configuration,
representing the optimal text generation under n-gram constraints.




In the Appendix, we introduce a pruning strategy 
to eliminate low-probability cases while maintaining a balance between low perplexity and model generalization. Additionally, we incorporate Laplace smoothing to enhance performance.  
The ground state is obtained via adiabatic quantum computing, variational methods, or Grover's algorithm, with results for the latter two presented in Appendix. 

\section{Conclusion\label{sec:Conclusion}}
We introduce a quantum generative framework grounded in reversible computing principles, whereby a  generative task is mapped to a constrained parent Hamiltonian problem.
By constructing local Hamiltonians whose ground state encodes the target data distribution and coupling them into a global Hamiltonian, solving for the ground state via adiabatic evolution, variational methods, or Grover algorithm yields the quantum states that enable image and text generation.
The highly degenerate ground-state manifold substantially lowers the computational overhead relative to generic quantum many-body problems.

This approach is conceptually distinct from neural network-based methods: rather than training a parametric model, we directly encode generative constraints into a physically motivated Hamiltonian. The framework inherently leverages key quantum resources—superposition and entanglement—to maintain global consistency across outputs and to represent the solution space in superposition, from which optimal configurations can be extracted efficiently. 

While the current formulation provides a foundational protocol, the limited number of qubits presents a key constraint. 
Future work will pursue two directions: resource-efficient Hamiltonian construction (e.g., tensor networks or neural-network ansatze) and optimized mappings between quantum states and outputs to enhance information density. These advances may yield practical quantum-inspired classical algorithms and near-term quantum implementations.

\begin{acknowledgments}
This work was supported by China Natural National Foundation of China (Grant. No.12075059  and No.T2241005)
\end{acknowledgments}
\newpage
\appendix

\section{\label{app:pruning}Pruning Strategy}
To suppress low‑probability components, we can introduce a local pruning operator that acts on sliding windows of the weight registers:
\begin{equation}
m^{\,l} \,| \{w_k\} \rangle = \begin{cases}
| \{w_k\} \rangle & \displaystyle\prod_{k\in W_l} w_k < \theta_l, \\[10pt]
0 & \text{otherwise},
\end{cases}
\end{equation}
where $W_l$ denotes the $l$-th window (with partial overlap between adjacent windows) and $\theta_l$ is a threshold.

\section{\label{app:smoothing}Smoothing Mechanism}
To enhance generalization, smoothing can be incorporated directly into the Hamiltonian. As an example, we show how Laplace (add‑one) smoothing is realized. The total Hamiltonian becomes
\begin{equation}
H_{\text{total}} = \sum_{k=1}^{T-n+1} \bigl(H_k + L_k\bigr) \;+\; \lambda M,
\end{equation}
where $L_k$ are smoothing operators and $\lambda M$ is the pruning term.

Each $L_k$ acts on the $k$-th weight register as
\begin{equation}
L_k \,| m \rangle _{\text{weight}} = \begin{cases}
-E_g \,| m \rangle _{\text{weight}}, & m = 0, \\[4pt]
0, & \text{otherwise}.
\end{cases}
\end{equation}
The negative energy $-E_g$ favors the zero‑weight state, effectively implementing the “add‑one” correction.

The combined Hamiltonian $H_k+L_k$ yields an enlarged ground‑state subspace
\begin{equation}
\mathcal{G} = \mathcal{G}_{H} \;\cup\; \mathcal{G}_{L},
\end{equation}
which explicitly reads
\begin{equation}
\begin{aligned}
\mathcal{G} = &
\underbrace{\Bigl\{\bigotimes_{j=1}^n | x_j \rangle_{t_j} \otimes | k \rangle \;\Big|\;
\vec{x} \in \mathcal{V},\; k=1,\dots,w \Bigr\}}_{\mathcal{G}_{H}} \\
& \cup \;
\underbrace{\Bigl\{\bigotimes_{j=1}^n | x_j \rangle_{t_j} \otimes | 0 \rangle \;\Big|\;
\forall\, \{x_j\} \Bigr\}}_{\mathcal{G}_{L}} .
\end{aligned}
\end{equation}
The first set contains the original weighted patterns, while the second set includes every n-gram with the smoothed (zero‑weight) label, thereby extending the model’s coverage to unseen token combinations.

\section{\label{app:text example}Examples of text generation}
To demonstrate our quantum text generation framework, we start from a dataset comprising 402 three-letter English words and utilize adjacent word pairs as 2-gram grammar to generate three-letter words in a 21 qubit system.

For each 3-letter word $x_1x_2x_3$, we establish two distinct bigram probability models:
\begin{equation}
\begin{aligned}
\mathcal{P}^{12} &= (x_1, x_2, P(x_2|x_1)) \\
\mathcal{P}^{23} &= (x_2, x_3, P(x_3|x_2))
\end{aligned}
\label{eq:bigram_models}
\end{equation}

Each transitional probability pattern $\mathcal{P}^{12}$ and $\mathcal{P}^{23}$
maps to an orthogonal state ensemble:

\begin{equation}
\mathcal{S}^{12} = \left\{ | x_1 \rangle_{t_1} | x_2 \rangle_{t_2}  | k \rangle_{w} \,\bigg|\, 
 k \in \{1,2,\ldots,w_1\} \right\} 
\label{eq:bigram_encode}
\end{equation}

\begin{equation}
\mathcal{S}^{23} = \left\{ | x_2 \rangle_{t_2} | x_3 \rangle_{t_3} | k \rangle_{w} \,\bigg|\, 
\, k \in \{1,2,\ldots,w_2\} \right\}
\label{eq:bigram_encode_s23}
\end{equation}
where the weight registers $w_1$ and $w_2$ store quantized transitional frequencies:
\begin{equation}
\begin{aligned}
w_{1} &= \lfloor C \cdot P(x_2|x_1) \rfloor \\
w_{2} &= \lfloor C \cdot P(x_3|x_2) \rfloor
\end{aligned}
\end{equation}

The 21-qubit system is allocated as:
\begin{equation}
\begin{aligned}
\text{Token1 Register} &: \text{qubits } 0-4 \\
\text{Token2 Register} &: \text{qubits } 5-9 \\
\text{Weight1 Register} &: \text{qubits } 10-12 \\
\text{Token3 Register} &: \text{qubits } 13-17 \\
\text{Weight2 Register} &: \text{qubits } 18-20
\end{aligned}
\label{eq:qubit_allocation}
\end{equation}

\begin{figure}[ht]
    \centering
    \includegraphics[width=\linewidth]{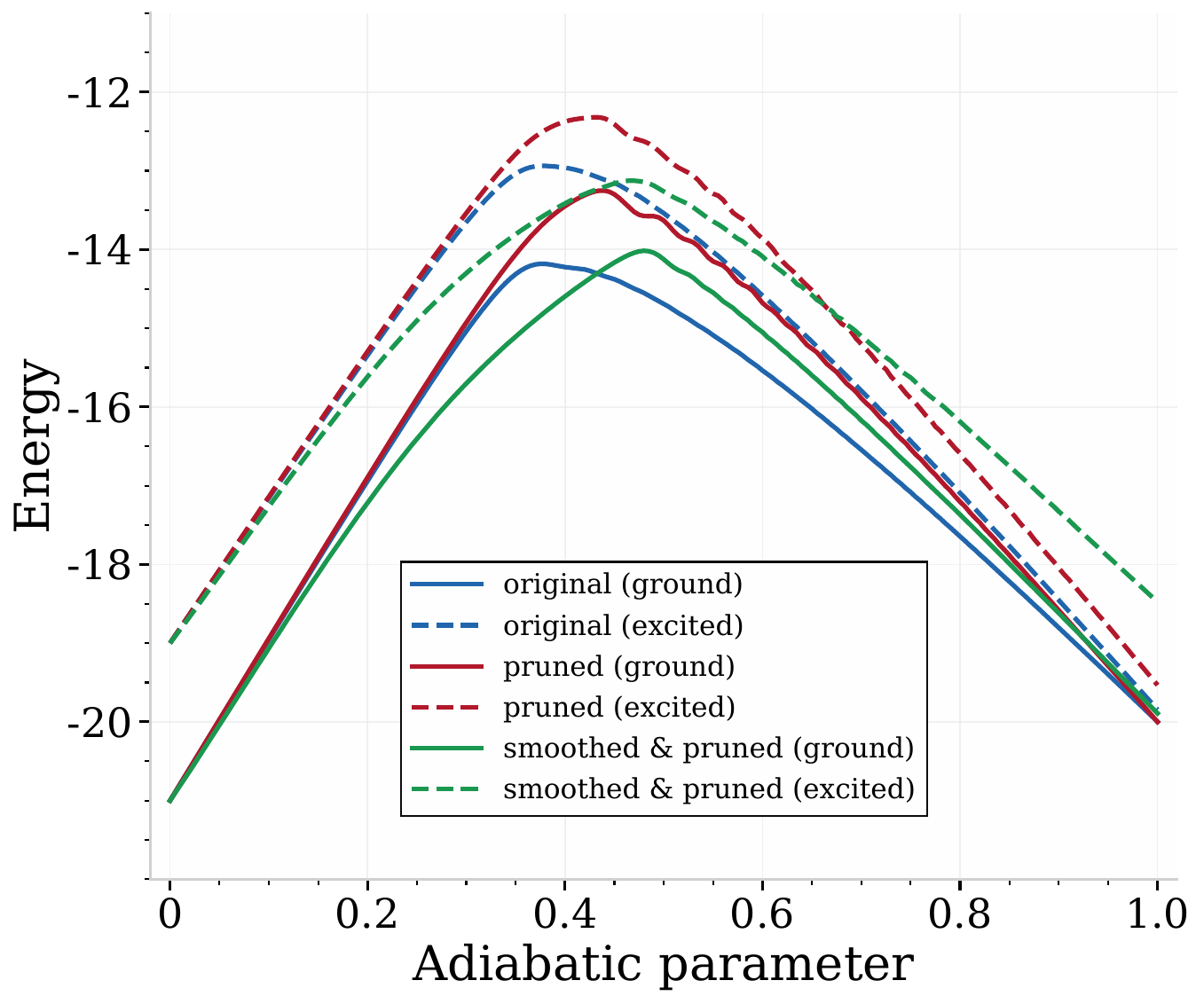}
    \caption{
        This figure shows the energy of ground and excited states under three protocols via adiabatic evolution: (i) Origin (no pruning/smoothing), (ii) Pruning-only, (iii) Pruned-smoothed.
        The Hamiltonian with a highly degenerate ground state manifold can still evolve to the ground state even for complex systems. Furthermore, excited states do not evolve into the ground state when the pruning strategy is applied.
    }\label{fig:word_ad}
\end{figure}

We integrate Laplace smoothing and quantum pruning through the modified Hamiltonian:

\begin{equation}
H_{\text{total}} = \sum_{k=1}^{2} (H_k + L_k) + \lambda M
\label{eq:modified_hamiltonian}
\end{equation}

where the operator domains are explicitly defined as:
\begin{itemize}
\item $H_1$ acts on the $(x_1, x_2, w_1)$ registers (qubits 0--4, 5--9, 10--12)
\item $H_2$ acts on the $(x_2, x_3, w_2)$ registers (qubits 5--9, 13--17, 18--20) 
\item smoothing operator $L_1$, $L_2$ act on the $w_1$, $w_2$ registers (qubits 10--12; 18--20)
\item Pruning operator $M$ operates on the combined weight registers $(w_1, w_2)$ (qubits 10--12, 18--20)
\end{itemize}
We can solve this using variational or adiabatic approaches, where Figure \ref{fig:word_ad} specifically displays the energy spectrum of annealing evolution.
\section{\label{app:Grover}Grover Algorithms}
\begin{figure}[ht]
    \centering
    \includegraphics[width=\linewidth]{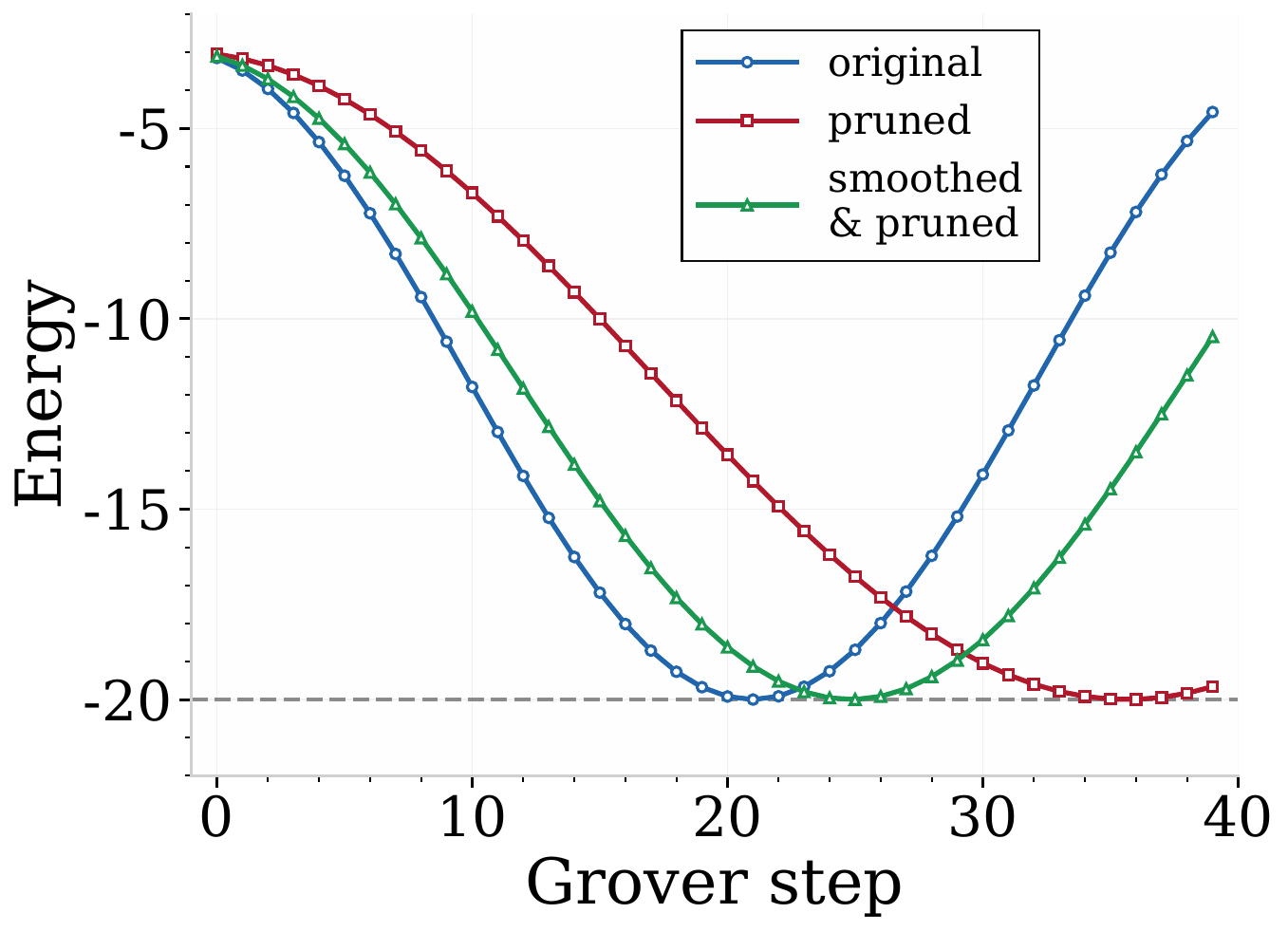}
    \caption{Energy expectation of three protocols: (i) Origin (no pruning/smoothing), (ii) Pruning-only, (iii) Pruned-smoothed. 
    We can get the ground state in 20--40 Grover iterations.The Grover algorithm inherently constructs an exact uniform superposition of all allowed patterns.
    }\label{fig:word_grover}
\end{figure}

Our generation task lends to a Grover formulation where:
\[
f(x_1x_2x_3) = 
\begin{cases}
1 & \text{iff } x_1x_2x_3 \in \mathcal{V}  \\
0 & \text{otherwise}
\end{cases}
\]
The verification oracle decomposes as:
\begin{equation}
\label{eq:grover decompose}
    f(x_1x_2x_3) = f_1(x_1x_2w_1) \land f_2(x_2x_3w_2) \land f_{\text{prune}}(w_1w_2)
\end{equation}
$f_1$ encodes all configurations of the first two letters with their weights,
which can be decomposed into two components:
\begin{equation}
f_1(x_1x_2w_1) = f_{H_1}(x_1x_2w_1) \oplus f_{L_1}(w_1)
\end{equation}
where the XOR ($\oplus$) enforces mutual exclusivity of constraints
due to $H_1$ and $L_1$ having orthogonal ground states.
The pattern verification operator implements quantum disjunction:
\begin{equation}
f_{H_1}(x_1x_2w_1) = \\
\bigvee_{k\in S_{12}} \underbrace{\mathbb{I}[x_1x_2w_1 \equiv k]}_{\text{Pattern detector}}
\end{equation}
where $\mathbb{I}[\cdot]$ is the indicator function
returning 1 iff $x_1x_2w_1$ exactly matches pattern $k$.
The $\mathbb{I}$ gate can be constructed through a combination of Pauli-X gates and a multi-controlled X gate.
The complete oracle implementation becomes:
\begin{equation}
\text{Oracle}(x_1x_2x_3w_1w_2) = O_1 \cdot O_2 \cdot O_p \cdot \text{AND} \\
\cdot O_p \cdot O_2 \cdot O_1
\end{equation}
Figure~\ref{fig:word_grover} quantifies the energy expectation after each Grover-iteration.


\providecommand{\noopsort}[1]{}\providecommand{\singleletter}[1]{#1}%
\end{document}